\documentclass[conference]{IEEEtran}
\usepackage[normalem]{ulem}

\usepackage{subfiles}
\usepackage{multirow,multicol,tabularx} 
\usepackage{booktabs}
\usepackage{subcaption}
\usepackage{amsmath}

\usepackage{relsize}
\usepackage{pifont}
\usepackage{amsmath}
\usepackage{subfig}
\usepackage{graphicx}
\usepackage[table]{xcolor}
\usepackage{tikz}
\def\biblio{\bibliographystyle{unsrt}\bibliography{\main/references}} 

\newcommand{\circled}[1]{\tikz[baseline=(char.base)]{
            \node[shape=circle,fill,inner sep=0pt] (char) {\textcolor{white}{#1}};}}

\renewcommand{\baselinestretch}{0.985} 
\begin{document}
\def\biblio{}
\noindent

\title{The Case for Persistent CXL switches}
\author{
  Khan Shaikhul Hadi\\
  \smaller University of Central Florida\\
  \smaller \emph{shaikhulhadi@ucf.edu}
  \and
  Naveed Ul Mustafa\\
  \smaller New Mexico State University\\
  \smaller \emph{num@nmsu.edu}
  \and
  Mark Heinrich\\
  \smaller University of Central Florida\\
  \smaller \emph{heinrich@ucf.edu}
  \and
  Yan Solihin\\
  \smaller University of Central Florida\\
  \smaller \emph{yan.solihin@ucf.edu}
}
\date{}
\maketitle

\begin{abstract}

Compute Express Link (CXL) switch allows memory extension via PCIe physical layer to address increasing demand for larger memory capacities in data centers. However, CXL attached memory introduces 170ns to 400ns memory latency. This becomes a significant performance bottleneck for applications that host data in persistent memory as all updates, after traversing the CXL switch, must reach persistent domain to ensure crash consistent updates. We make a case for persistent CXL switch to persist updates as soon as they reach the switch and hence significantly reduce latency of persisting data. To enable this, we presented a system independent persistent buffer (PB) design that ensures data persistency at CXL switch. Our PB design provides 12\% speedup, on average, over volatile CXL switch. Our \textit{read forwarding} optimization improves speedup to 15\%.

\end{abstract}

\section{Introduction} \label{introduction}
The emergence of Compute Express Link (CXL) interconnect technology is a significant event that addresses the ever increasing demand for higher memory capacity in data center servers. With CXL, memory devices can be attached to the PCIe ports, accessible by the CPU via loads/stores instructions~\cite{cxl_introduction2024}, allowing {\em memory expansion} over traditional (local) memory. Due to its packet-based protocol, CXL enables various types of memories to be attached, including volatile (DRAM) and byte-addressable non-volatile memory (Optane Pmem, memory-semantic SSD, etc.). From beginning, the CXL protocol was designed to be compatible with persistent memory (PM) devices, with features to manage persistency (e.g. global persistent flush)~\cite{cxl_support_pm}. This is important because PM devices such as memory-semantic SSDs are projected to be much denser, hence provide a much larger capacity, than traditional DRAMs. Furthermore, many applications including databases and key-value stores, can benefit from the persistency in addition to the capacity~\cite{PM_Architecture_book}.  

While CXL attached memory incurs significant lower latency than network-attached memory, it still incurs significantly higher latency (e.g. 170ns$\sim$400ns\cite{memory_disaggregation2023}), $2-3\times$ higher compared to the local main memory access, due to CXL switch traversal. 

\begin{figure}[!ht]
    \centering
    \includegraphics[scale=0.5]{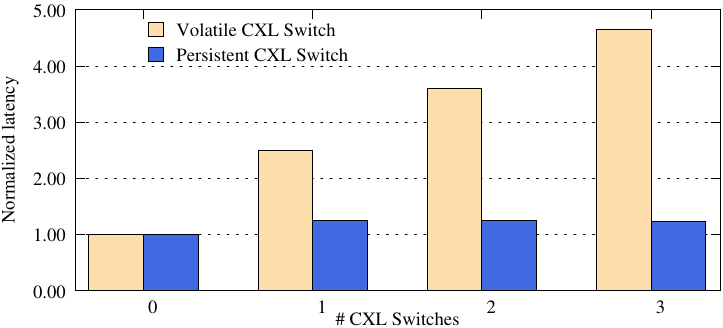}
    \caption{Normalized latency of persist operations in FFT.}

    \label{fig:persistent_cxl_persist_latency_reduction}
\end{figure}

Persistency-aware applications need to manage crash consistency, where upon a crash (including power loss), the state in PM is consistent such that it allows computation to recover from the crash. An application orchestrates crash consistency by explicitly specifying {\em persists}, i.e. stores that must reach persistency before the application can execute other instructions~\cite{rudoff2017persistent}, typically achieved by using cache line flush and memory fence instructions. With CXL-attached PM devices, each persist must traverse the CXL network, exposing the entire round trip latency to PM. Figure~\ref{fig:persistent_cxl_persist_latency_reduction} illustrates the magnitude of the problem, showing the persist latencies (yellow bars) for FFT from Splash-4 benchmark suite~\cite{splash4}, normalized to a system without CXL switch. The x-axis shows the number of CXL switches a request must traverse to reach PM where $n=0$ indicates that PM is local main memory. For PM as memory expander \cite{samsung_memory_expanders} connected via a single CXL switch, data hops a single CXL switch to reach PM, resulting in 2.5$\times$ persist latency. For larger CXL systems~\cite{cxl_3_introduction_2022}, the persist may traverse multiple CXL switches before reaching PM, resulting in an exponential increase in persist latency. Overall, the figure shows that the persist delays are very substantial and will get worse fast as the system size increases. 

To reduce persist latencies, one may consider techniques from prior works like {eADR}~\cite{eADR} and BBB~\cite{bbb} where the CPU chip is equipped with battery backing such that data can be persisted at the CPU cache. Unfortunately, such a {\em processor centric} approach has two major drawbacks: it requires substantial changes to the CPU architecture and requires batteries, and it assumes PM to be attached as local memory. Furthermore, CXL-attached PM has a wide range of memory access latency and bandwidth characteristics~\cite{demystyfy_cxl2023}. Processor-side solutions must provision the correct battery capacity and buffer sizes to keep the cost low, but the uncertainty of latencies to CXL-attached PM devices may force battery and buffer over-provisioning, making them excessively costly. 
In this paper, we propose a novel {\em memory centric} approach, where CXL switches are re-designed to enable accepting persists. Our approach maintains the CPU-agnostic characteristics of CXL protocol and switches. The potential benefits of this approach is illustrated in the blue bars in Figure~\ref{fig:persistent_cxl_persist_latency_reduction}, where a persist only needs to reach the first CXL switch to complete. Overall, the figure shows that persist latency bottlenecks can be largely mitigated if persists can complete at the first CXL switch.

In this work, we propose a persistent CXL Switch (PCS) by adding a Persistent Buffer (PB) to the switch. Our PB is system-independent as it requires no processor-side modifications. Designing PCS comes with two main challenges. \textit{Correctness challenge} requires that a) any read request for persistent data must be supplied with latest copy which may reside in a PCS in the network or in PM and b) updates are persisted in PM in the program-order. \textit{Performance Challenge} requires that no extra latency should be incurred for unrelated packets (e.g., volatile data) and unrelated CXL protocols (i.e., CXL.io and CXL.cache). To address these challenges, our design of PCS is capable of identifying persistency-related CXL packets which enables it to ensure data correctness and avoiding extra latency by treating those packets differently. Our design also introduces Read Forwarding (RF) optimization that supplies data from PB, when possible, instead of PM to any subsequent read requests, lowering the read-latency of persistent data.



Our proposed design of PCS offers multiple benefits. First, it significantly reduces latency of persist operations. Second, our PB design and its operation does not require any changes in the CXL protocol. Third, our design of PB adds persistency to CXL switch independent of CXL architecture. In other words, our design can be applied on CXL switch from different vendors. Finally, as our design of PB does not expect any modifications to the CPU it could be compatible with any \textit{processor centric} optimization.



To summarize, our work makes following contribution:
\begin{itemize}
    \item We propose a novel memory-centric persist approach to manage CXL-attached PM. 
    \item We present the design of Persistent CXL Switch (PCS), adding a Persistent Buffer (PB) to the switch, addresses the correctness and performance challenges.
    \item We evaluated PCS using Splash-4 parallel benchmark suite. Our evaluation shows that PCS significantly reduces the latency for persist operation (up to 55\%) resulting in 12\% speedup, on average, of the evaluated workloads.
    \item We propose Read Forwarding (RF) optimization that improves speedup to 15\% on average.
\end{itemize}






\section{Background}

\subsection{Compute Express Link}
Compute Express Link (CXL) is an open industry-standard interconnect protocol developed to facilitate high-bandwidth, low-latency communication between processors (called hosts) and devices such as memory devices and accelerators. CXL enables memory expansion, memory disaggregation, and memory pooling~\cite{cxl_enable_datacentric_computing2023, cxl_introduction2024, cxl_enahnced_function2023}. CXL protocol consists of CXL.io, CXL.cache and CXL.mem which are dynamical multiplexed over the PCIe physical layer. CXL.cache is used to maintain coherence in shared data, CXL.mem is used to read write 64-byte cache line data from memory and CXL.io is based on PCIe mechanisms for I/O communication\cite{cxl_enahnced_function2023}. CXL devices can be one of three types: Type 1 device can access and cache host memory, Type 2 could coherently share their own memory with host and Type 3 devices are only load/store accessible (i.e. CXL-attached memory). Since CXL v1.0, coherence and memory semantics have been supported~\cite{cxl_introduction2024, cxl_standard_datacenter2022} for devices including PM (e.g. memory semantic SSD~\cite{samsung_memory_semantinc_ssd} and Optane Pmem~\cite{optane_brief,kioxia}). This work focuses on CXL.mem with PM devices attached.

\subsection{Persistent Programming Model}

With PM, applications may host persistent data in PM. They need to manage crash consistency,  a property where upon a crash (including power loss), the memory state of PM is consistent such that it allows computation to recover from the crash. An application orchestrates crash consistency by explicitly specifying {\em persists}~\cite{rudoff2017persistent}, which are stores that must reach durability before other memory accesses can proceed. Persists are supported differently with different instruction sets. For example, Intel x86 provides \textit{clflush, clflushopt \& clwb} instructions to perform cache line flush or write-back and \textit{mfence, sfence} to ensure flushes reach the persistent at that point~\cite{intel64manual}. ARM ISA provides a combination of \textit{DC CVAP, DC CVADP} for write-back and \textit{DSB} as a fence~\cite{arm_manual}. Those instructions may be used in a library (e.g., Intel PMDK\cite{pmemio}) or directly in the application program to enforce persists and achieve crash consistency. Fundamentally, a persist makes program wait until data is written to PM before continuing, therefore persist latency is critical to performance. 


\section{Motivation}

\begin{figure}[!ht]
    \centering
    \includegraphics[width=\linewidth]{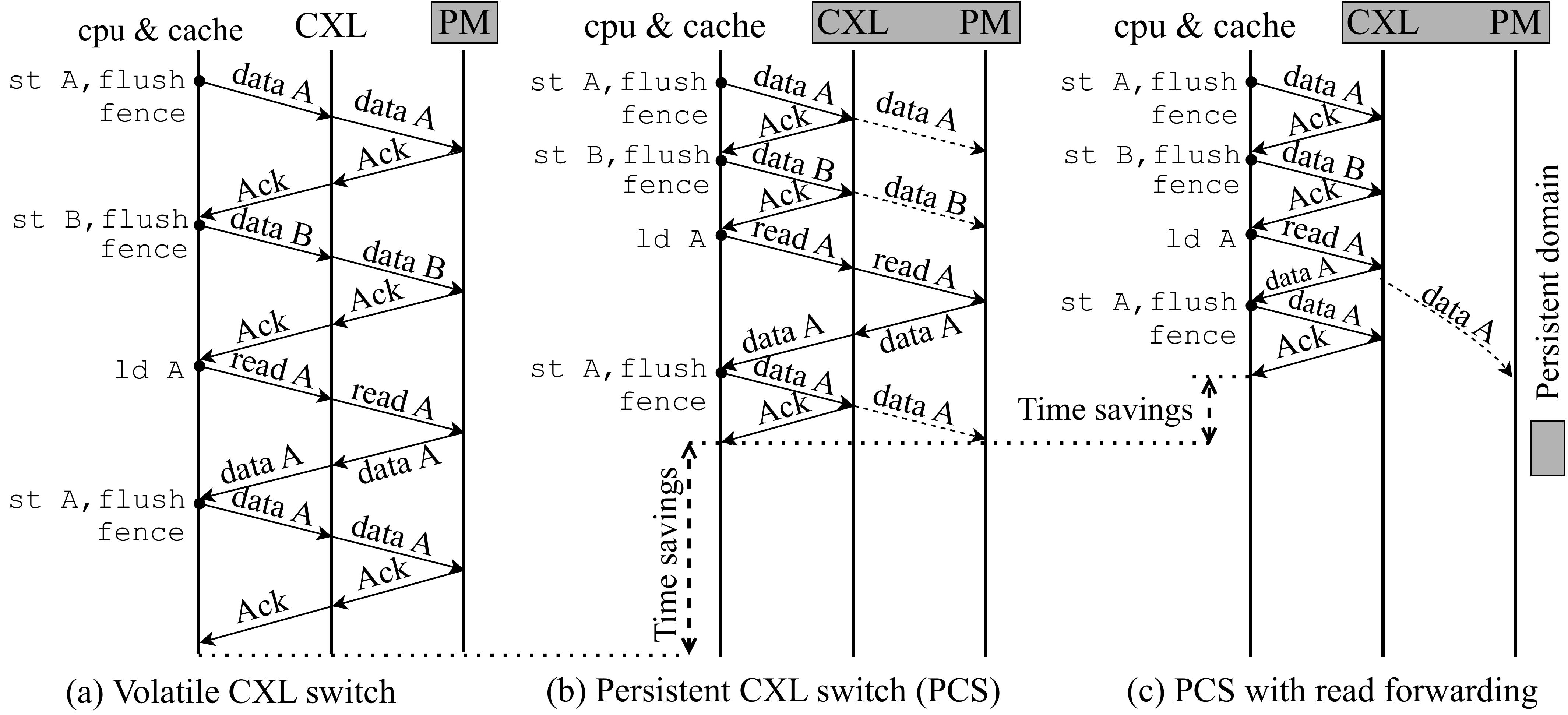}
    \caption{Potential time savings for persistent CXL switch(PCS)}

    \label{fig:pcs_motivation}
\end{figure}

Here, we will illustrate the performance problem with persists in a CXL-attached PM system, and how our approach may solve it. First, we define \textit{Persistent Domain} (PD) as components that can accept a persist such that durability is guaranteed, i.e. its value is not lost on a crash. A persist is complete when it is accepted in the PD. Figure \ref{fig:pcs_motivation}(a) the persist timing in a traditional CXL system. Suppose that the CPU wants to persist a store to address A (using a pair of flush and fence), a store to B, load from A, and persist a store to address A for the second time. In a current system, each persist requires sending data to the CXL switch and PM, and for the acknowledgment to come back to the CPU before the fence can be retired from the processor pipeline, after which the next persist can execute. Thus, four full memory round trips occur.  

Figure \ref{fig:pcs_motivation}(b) illustrates the benefit our our approach. The CXL switch is made persistent hence the persistent domain expands to include CXL switches. This necessitates the CXL switch persist buffer to use non-volatile cells or have a small battery backup that has enough spare energy to drain buffers to the PM. When a persist A reaches the first persistent switch (PCS), PCS replies with an acknowledgment to the CPU that the persist has completed, while in the background the PCS drains data to the PM. By overlapping the acknowledgment and sending data to the PM, we reduce the critical path delay for the three persists, resulting in substantial latency savings. The same time saving occurs for the persist of B and A. However, the load to A will still go to the PM because the block was already drained to PM earlier.  

Note that the remaining roundtrip to memory (for load A) can be eliminated as well, if instead of draining data immediately to the PM, we retain them in the CXL switch as long as possible. As shown in figure \ref{fig:pcs_motivation}(c), if we keep data in the PCS, the PCS can just provide block A to the load instruction. It will result in time saving as the read avoids the full roundtrip to PM. 

\section{Design Principle}
 
\subsection{Correctness Criteria }

To ensure data correctness and crash consistency, PCS design must meet following criteria: 

\paragraph{Write-Read order} An application may update persistent data, followed by subsequent reads of the same data. Since PCS belong to PD, its design should include a mechanism to identify the latest copy of the persistent data, whether it resides in PCS or in PM. Once located, the data should be supplied to the read request from the appropriate source.
    

\paragraph{Write order} An application may update the same persistent data multiple times during its execution. Design of PCS must ensure that these updates are persisted in the same order as they enter PD. Otherwise, an older persistent update may overwrite the younger one violating data correctness.


\paragraph{Crash consistency} To ensure crash consistency, once data persistency is confirmed by any persistent structure, data must not leave persistent domain ( always a persistent copy of the latest version of the data exit in the persistent domain). if crash happen while data is out of persistent domain, data will be lost or get corrupted upon reboot.PCS must ensure there will always be a persistent copy of the the data in the  persistent domain before sends acknowledgment of data persistency.

 \subsection{ Assumptions}
   We assume a system where CXL switch is used to expand memory capability as shown in Fig \ref{fig:system_design_for_PB}(a). DRAM is volatile main memory and PM is CXL switch attached extended memory. All heap memory are mapped to PM and non-heap memory is considered volatile and mapped to DRAM. Program use \textit{clflush} instruction to flush all copy from volatile domain to PD for durability \textit{mfence} to enforce persist barrier.  For CXL switch design, as there is no publicly available information regarding how actual cxl supported hardware are designed, we assumed a simple four stage pipeline router design with cxl switch latency shown in Pond \cite{pond_2023}. However, our proposed design is independent of switch design and will work on any other variation of switch hardware design as long as certain criteria is met (see \ref{selector_design}). No packet loss or data loss is assumed during packet transmission via CXL switch. Routing packets is considered volatile thus upon crash packet loss is assumed.  To persist data in the buffer, we assume our buffer use persistent memory technology. However, our design also work for battery backed system as long as enough backup power is guaranteed to write back data in PM. To write-back data in the PM, persist buffer require metadata to create cxl protocol packet. Detail of required meta data dependent on specific cxl hardware design, thus not a focus of our work. To demonstrate our idea, we assume 16 Byte generic slot \cite{cxl_specification2023} information related to the request from cxl packet header as required metadata. 

\subsection{Design Challenges}
\subsubsection*{Performance challenge}
Detecting and processing a CXL packet for reading and persisting data in the persist buffer (PB) requires time. However, PCS must also handle irrelevant packets (e.g., CXL.io, CXL.cache) without interfering with routing. To avoid added latency, the PCS design needs a parallel mechanism to detect persistency-related packets. When detected, such packets must be removed from the critical path to prevent stalls while the controller persists data or waits for resources. Adding latency to all packet routing could severely impact overall system bandwidth and become a performance bottleneck.

\subsubsection*{Correctness challenge}
With PB in PCS, the latest data version may reside in PB, PM, or both. The design must detect the latest version and ensure reads return most recent copy. During data draining, the write-back request must reach PM before read request to maintain write-read order. PCS must confirm data is persisted in PM before considering drain complete to ensure crash consistency. When multiple data versions arrive, PCS must enforce write order to prevent older versions from overwriting newer ones.

\subsection{ Optimization: Read Forwarding (RF)}
\label{rf_optimization}
Workload with high temporal locality  may request read within short interval after flushing data to PD. Forwarding data from PCS to the CPU  save roundtrip time from CXL switch to PM (figure \ref{fig:pcs_motivation}(c)). Read forwarding (RF) optimization aims to keep data at PCS as long as possible to respond read with correct data. Workload with high temporal locality on write operation  persist update on same address frequently. Considering PCS keeps data to itself for read forwarding,  it could update data with new version without flushing old data to PM. We address this as write coalescing. Write coalescing could reduce number of write on PM, effectively increasing life-cycle of PM.  Reduce data flush to PM may reduce network traffic in the downstream. 
 
\begin{figure*}[!h]
    \centering
    \includegraphics[width=0.95\textwidth]{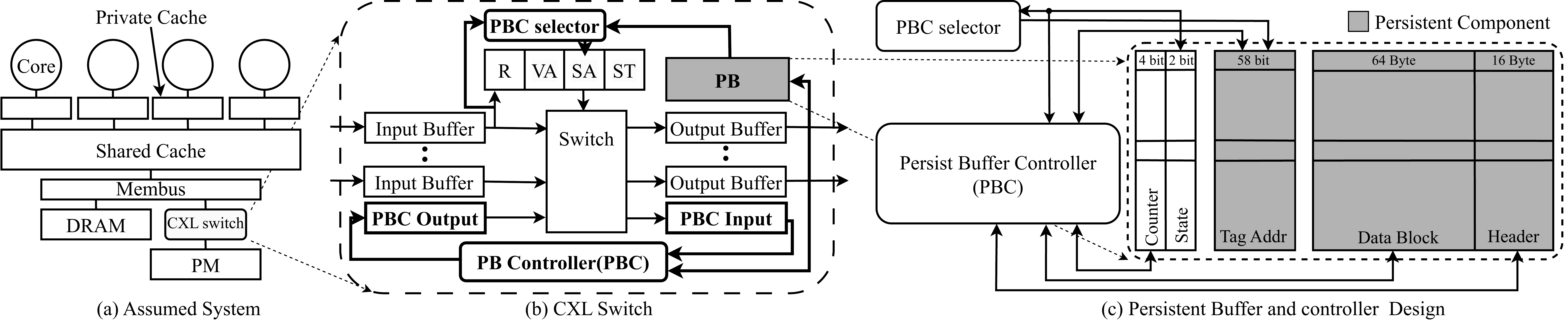}
    \caption{Persistent Buffer Design for Persistent CXL switch.}
    \label{fig:system_design_for_PB}
\end{figure*}

\begin{figure*}[h]
    \centering
    \includegraphics[width=0.95\textwidth]{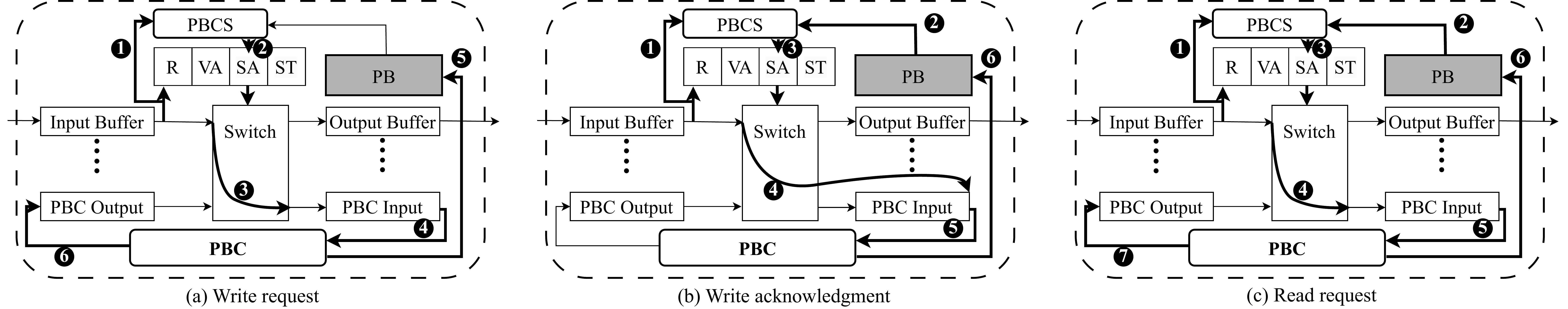}
    \caption{ {Workflow of PB for read,write and acknowledgment packet.} }
    \label{fig:pb_work_flow}
\end{figure*}

 \section{Persistent CXL Switch Design}

\subsection{Persistent Buffer (PB)}
PB is a persistent technology based fully associative cache like structure.  PB consist of three  table: \textit{Data Table(DT)}, \textit{Tag Address Table (TAT)} and \textit{State Table (ST)}. Each PB entry (PBE) consist of one TAT entry and associated entry from DT and ST. As shown in Figure \ref{fig:system_design_for_PB}(c), Each DT entry consist of 64 Byte data block and 16 Byte header (metadata). Each TAT entry consist of 58bit tag address for 64bit address. As we save data with cache line granularity,  6bit LSB of 64bit address is not needed. Each ST entry consist of 2 bit state and 4 bit counter for LRU replacement policy. PBE could have 3 states: \textit{Dirty, Drain \&  Empty}. \textit{Dirty} indicate data at PBE is the latest \& only version and next persistent structure (i.e. PM) have stale copy. \textit{Drain} indicate a copy of the data is sent to next persist structure to persist. \textit{Empty} indicate data is persisted in next persistent structure and this entry could be treated as empty entry to serve new persist request.

\subsection{Persist Buffer Controller (PBC)} 
PBC is primary logic circuit for this version of PCS design. PBC is connected to two buffer channel: PBC Input (PI) Buffer and PBC Output (PO) Buffer; similar to input buffer and output buffer of the switch (Figure \ref{fig:system_design_for_PB}(b)). PI buffer have additional feature that it prioritize \textit{write acknowledgment} response over read/write request to prevent potential deadlock(see section \ref{acknowledgement_packet}) in PBC. Switch route relevant packets and place at the end of PI buffer. PBC pick one packet at a time from the front of the buffer to ensure sequence is maintained for data correctness.  

To persist a new data block, PBC finds an \textit{Empty} PBE and update tag address , data block, header information, PBE state \& reset counter. If no \textit{Empty} PBE found, PBC chooses a victim PBE with state \textit{Dirty} and start draining data to PM. To drain a PBE, PBC change entry state from \textit{Dirty} to \textit{Drain}, read tag address, data block and use header information to create a CXL packet to insert it at PO buffer for the switch to eventually route it to it's destination (i.e. PM).

\subsection{Persist Buffer Controller Selector (PBCS)} 
\label{selector_design}
PBCS is responsible for analyzing packet header to decide if it is a relevant packet and needs to send to PBC. PBCS works parallel of routing pipeline so that it does not slow down routing process. PBCS is allowed to read state of a PBE for supplied tag address. Based on the state information along with packet header information, PBCS decides if a packet needs to go to PBC or not and sends this information to  switch allocator(SA) as input signal. It is crucial that PBCS completes it's decision before SA stage starts as SA will prioritize PBCS output over routing algorithm and route the packet to PI buffer if PBCS dictates. As long as this condition is maintained, regardless of how CXL switch hardware is implemented by different vendors, our PB design will work. When A packet sent of PI buffer, it effectively being removed from critical path of routing which guarantees uninterrupted routing of irrelevant packets regardless how long PBC will take to process this request.
\subsection{Working principle}
\subsubsection{Write request}
In Figure \ref{fig:pb_work_flow}(a), when a write request reaches  front of the input buffer, PBCS start analyzing it's opcode and if address mapped to PM (\circled{1}). If condition is met, sends signal to SA (\circled{2}) to route packet to PI buffer (\circled{3}). Eventually PBC retrieve this packet from PI buffer (\circled{4}) and look for an \textit{Empty} PBE in PB to persist data (\circled{5}). For multiple \textit{Empty}
entry, LRU policy is followed. If no \textit{Empty} entry found, if available, a \textit{Dirty} PBE using LRU policy will be chosen as victim entry. PBC change victim's state to \textit{Dirty} and start draining to PM ( preexisting \textit{Drain} PBEs are not considered for victim selection ). PBC stall following requests at PI  buffer until any PBE entry become \textit{Empty} to persist currently  waiting request at the front of PI buffer.  If not stalled, PBC may end up serving following read/write request and violate correction criteria.  If all PBE state is \textit{Drain}, PBC stall request processing  until any entry become \textit{Empty}. Upon data is persisted, PBC generate \textit{write acknowledgment}  response  for the source and place it in the PO buffer (\circled{6}).

For PB design, after replying with \textit{write acknowledgment}, PBC start draining the persist entry to free resource  as soon as possible for new request. This reduces risk of stall due to shortage of \textit{Empty} PBE . For RF, no entry is drained until \textit{Dirty} entry count  reaches a certain \textit{drain threshold} ( i.e. 80\% of PBE count). When \textit{Dirty} entry count cross the threshold,  PBC start draining PBE using LRU policy until \textit{Dirty} entry count reduces to a \textit{preset} value (i.e. 60\% of PBE count). Keeping \textit{preset } number of PBE as \textit{Drity} allow read forwarding and write coalescing from PCS. However, As PBE always wants to keep more than \textit{preset} number of  PBEs \textit{Dirty} (less \textit{Emptry} PBE), probability of stall due to shortage of \textit{Empty} PBE increase with higher \textit{preset} value.

\subsubsection{Write Acknowledgment}
\label{acknowledgement_packet}
After successful write, PM sends \textit{write acknowledgment} to confirm that data is persisted. As shown in Figure \ref{fig:pb_work_flow}(b), when PBCS detects this packet (\circled{1}), it check PB for \textit{Drain} PBE of target tag address (\circled{2}). If found, PBCS assumes PBC  is the intended receiver and signal SA to route this packet to PI buffer (\circled{3}) and PI buffer place it at front of all existing write requests (\circled{4}) at PI buffer to avoid deadlock. If \textit{write acknowledgment} response wait behind write request while PBC is waiting for this acknowledgment to \textit{Empty} PBE  from \textit{Drain} state to process write request, this results in  deadlock. Upon receiving acknowledgment response, PBC will update PBE state from \textit{Drain} to \textit{Empty} (\circled{6}).

\subsubsection{Read request}
\label{read_process_step}
As shown in figure \ref{fig:pb_work_flow}(c), When PBCS detects a read request (\circled{1}), it checks PB for associated PBE (\circled{2}). If no PBE or \textit{Empty} PBE is found, PBCS does not interfere with switch allocator from routing it to PM. For \textit{Empty} PBE, it is likely to be replaced by existing write request at PI buffer.  If state is \textit{Dirty} or \textit{Drain} , PBCS will inform SA (\circled{3}) to route request to PI buffer (\circled{4}). For \textit{Dirty} PBE, latest version only exist in PB. Eventually PBC process the read request (\circled{5}), read data from PB if exist (\circled{6}) and place read response at PO buffer (\circled{7}) to forward it to the requester. If data not found at PB (i.e. already drained \& replaced), it place the original request at PO buffer (\circled{7}) to route to PM. For PB scheme, we start draining a data block as soon as it is persisted in PB. Thus \textit{Dirty} PBE is never detected for PB scheme. If  \textit{Drain} PBE is detected at PBCS, it is possible that this PBE updated to \textit{Empty} and replaced by another write request before this read request could reach front of PI buffer while latest version resides in PM. However, routing read to PM may violate \textit{write-read order} correctness criteria as data may be at PO buffer while read is routed to PM. For correctness, read request is routed to PI buffer so that when PBC inject this request in PO buffer (\circled{7}) it  automatically enforced write read order. If this occurs too frequently, this will increase read response latency significantly resulting performance loss. 

\subsubsection{Crash Recovery}
As routing is volatile and packet will be lost upon crash, PBC never \textit{Empty} an PBE until it receives \textit{write acknowledgment} from PM. This ensures that PB always have the latest version of data in PB for tag addresses that exist in TAT and If PB does not have the tag address in the TAT, latest version is guaranteed to be in the PM.   If crash happen, after reboot, PBC will assume all entry is \textit{Dirty} and drain each and every entry of PBE. As it is guaranteed to have latest version in PB, it ensure that all copy of data in PM is the latest version by writing every PBE back to PM. 


\section{Experimental Setup}
\noindent
\subsubsection*{ Simulation Environment} We used gem5 \cite{gem52020,gem52011} (version 23.0.0.1), an event-accurate simulator to evaluate our work. We run our simulation on System Emulation(SE) mode with SimpleSwitchableProcessor() configuration (ATOMIC \& O3). Simulated system is detailed in Table \ref{tab:hardware_parameter}. To simulate CXL switch, we designed a 4 stage pipeline switch with latency profile mentioned in pond\cite{pond_2023}. To calculate access latency  for Tag Address Table entry and Data Table entry of PB we used cacti\cite{cacti1996} with 22nm configuration. 

\begin{table}[h!]
\centering
\caption{ {Simulation Environment}}
\begin{tabularx}{\linewidth}{|l|>{\arraybackslash}X|}
\hline
    Processor &  8-core OoO SimpleCore( 4 GHz each), 64-bit X86 ISA\\
    \hline
    \multirow{4}*{Cache}& 64-bytes block, inclusive, LRU replacement policy \\\cline{2-2}
                        & Private L1 cache:32KB, 8-way, 1 cycle tag latency\\\cline{2-2}
                        & TLB cache: 8KB, 4-way, 1 cycle tag latency \\\cline{2-2}
                        & Shared L2: 256KB, 8-way, 10 cycle tag latency  \\
    \hline
    \multirow{2}*{CXL switch}  & x16 lane, 30GB per lane,68Byte flit \\\cline{2-2}
                            & 4 stage pipeline \\

    \hline
    \multirow{3}*{Persist Buffer}  & 16 entry, LRU replacement policy \\\cline{2-2}
                            & 0.388ns tag access latency \\\cline{2-2}
                            & 0.785ns data access latency \\

    \hline
    \multirow{1}*{Total Memory}    & 6 GiB \\
    \hline
    \multirow{1}*{DRAM}    & SingleChannelDDR4\_2400 \\
    \hline
    \multirow{1}*{NVM}     & 100ns read latency , 200ns write latency\\
    \hline
\end{tabularx}
\label{tab:hardware_parameter}
\end{table}

\noindent 
\subsubsection*{ Evaluated Workloads}
\label{workload_list} 
We used the Splash-4 parallel programming benchmark\cite{splash4} and adopt efficient checkpointing \cite{efficient_checkpointing2017} for crash consistency of the workloads in persistent memory system. Table \ref{tab:benchmark} show custom  input parameters used for evaluation . We used ATOMIC cpu to speedup simulation and warm up the cache and switch to O3 cpu to collect statistics only for parallel region of interest (ROI). 
We run the workloads up-to 100,000  write operation to PM assuming all heap memory is allocated to PM. \textit{FFT, Radiocity \& Volrend\_npl} complete it's parallel region execution before exit requirement meet. Workloads that did not run properly for all scheme are omitted.

\begin{table}[h!]
\centering
\caption{Splash-4 Benchmark Suite}
\begin{tabularx}{\linewidth}{|l|>{\centering\arraybackslash}X||l|>{\centering\arraybackslash}X|}
\hline
    \rowcolor{lightgray} Benchmark & Input & Benchmark & Input\\
    \hline \hline
    cholesky       &  tk18.O           &    FFT            &  -m12 -l6         \\ \hline
    Lu\_cont       &  -n128            &   Lu\_non    &     -n128         \\ \hline
    Volrend\_npl    & head\-scaleddown2     & Raytrace     & teapot.env        \\ \hline
    Radiocity       & \multicolumn{3}{|c|}{-ae 5000 -bf 0.1 -en 0.05 -batch  }      \\ \hline  
\end{tabularx}
\label{tab:benchmark}
\end{table}

\noindent
\subsubsection*{Scheme Comparison}

We compare persist buffer (PB) based PCS scheme against system with volatile CXL switch that router all request to PM. We address volatile scheme as no persist buffer (NoPB) scheme. We also evaluated persist buffer with read forwarding (PB\_RF) optimization (see \ref{rf_optimization}).


\section{Evaluation}

\begin{figure}[!ht]
    \centering
    \includegraphics[width=\linewidth]{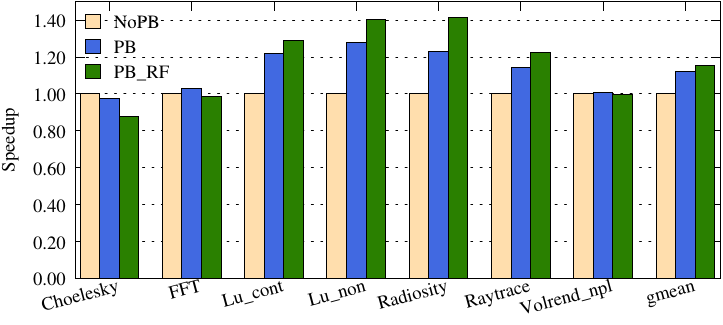}
    \caption{Speedup of PB and PB\_RF over NoPB (higher is better).}

    \label{fig:splash_speedup_evaluation}
\end{figure}

Figure \ref{fig:splash_speedup_evaluation} shows PB and PB\_RF achieve 12\% and 15\% average speedup over NoPB, respectively, for eight-threaded workload. The speedup is significant considering design simplicity and low hardware cost.
For \textit{Radiosity} and \textit{Lu\_non}, PB provides more than 20\% speedup where PB\_RF provides 40\% speedup. However \textit{Choelesky} suffers 3\% slowdown for PB and 13\% slowdown for PB\_RF. 

\begin{figure}[h]
    \centering
    \centering
    \begin{subfigure}[h]{0.24\textwidth}
        \includegraphics[width=1\linewidth]{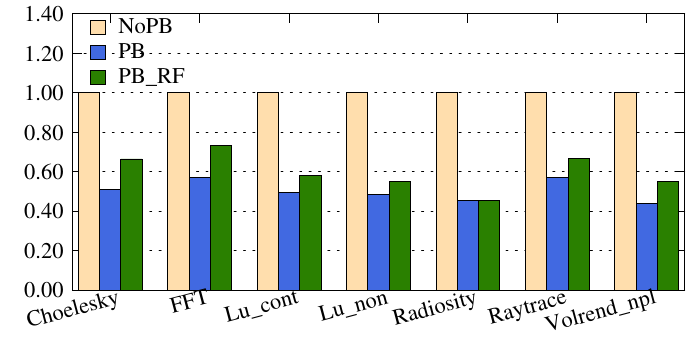}
        \caption{Normalized persist latency}
        \label{fig:persist_latency_evaluation}
    \end{subfigure}
    \begin{subfigure}[h]{0.24\textwidth}
        \includegraphics[width=1\linewidth]{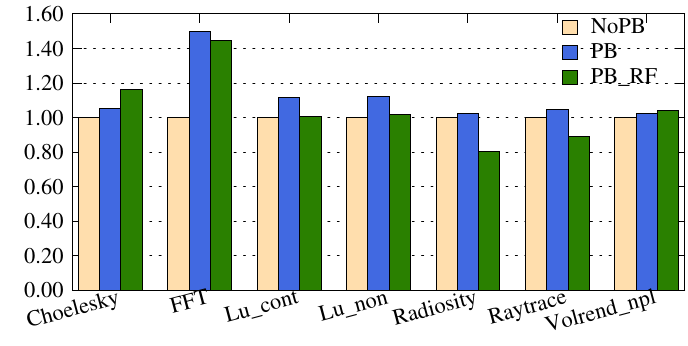}
        \caption{Normalized read latency}
        \label{fig:read_latecny_evaluation}
    \end{subfigure}

    \caption{Latency of persist and read operations (from LLC).}
    \label{fig:persist_and_read_latency}
\end{figure}



To assess the impact of PB and PB\_RF on different workloads, we examine persist and read operation latencies. As shown in Figure \ref{fig:persist_and_read_latency}(a) PB reduces persist latency by 43\% to 56\%, achieving the goal of PCS outlined in Section I. However, in PB\_RF, persist latency increases for most workloads, with varying rates. For \textit{Radiocity}, latency remains the same, while for \textit{Choelesky}, it rises from 51\% to 66\% of NoPB. A similar increase in persist latency is observed for \textit{FFT}. While PCS reduces persist latency as expected, its effect on overall workload performance is not straightforward (Figure \ref{fig:splash_speedup_evaluation}).





Figure \ref{fig:persist_and_read_latency}(b) explains why persist latency reduction doesn’t always lead to proportional performance improvement. Introducing PB increases read latency for all cases, as some reads are rerouted through the PI buffer (Section V-D3). Although the increase in read latency(2.5\%-12\%) is modest, workloads heavily reliant on read operations, like \textit{Choelesky} (which has a 2.5\% slowdown despite a 49\% reduction in persist latency), experience overall performance degradation. \textit{FFT}, however, does not suffer a performance loss (a 3\% speedup despite a 50\% increase in read latency). PB\_RF reduces read latency compared to PB for most workloads, except \textit{Choelesky} and \textit{Volrend\_npl}. For \textit{Choelesky}, PB\_RF increases read latency by 16\% over NoPB, causing a 13\% overall slowdown. Although FFT sees a reduction in read latency (from 50\% to 44\%), it still experiences a 2\% slowdown due to increased persist latency.

\begin{figure}[h]
    \centering
    \centering
    \begin{subfigure}[h]{0.24\textwidth}
        \includegraphics[width=1\linewidth]{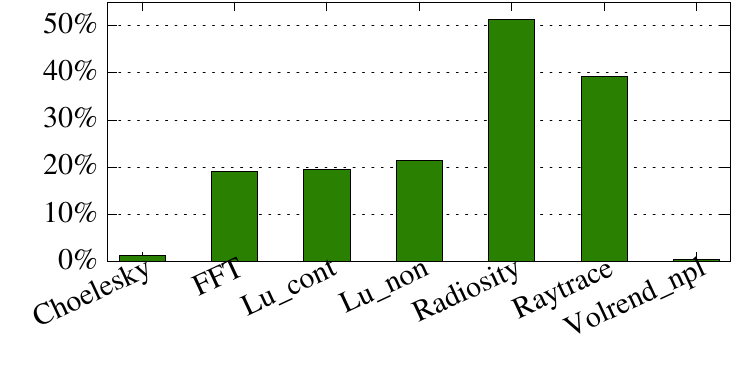}
        \caption{}
        \label{fig:read_hit_rate}
    \end{subfigure}
    \begin{subfigure}[h]{0.24\textwidth}
        \includegraphics[width=1\linewidth]{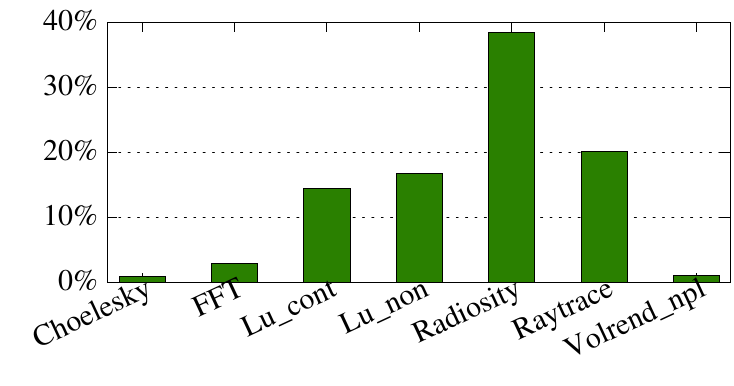}
        \caption{}
        \label{fig:write_coalescing_percentage}
    \end{subfigure}

    \caption{Read hit (a) and write coalescing rate (b) for RB\_RF.}
    \label{fig:pb_hit_rate}
\end{figure}

In Figure \ref{fig:pb_hit_rate}(a), we observe very high read hit rate at  {persist buffer} for \textit{Radiocity} (51\%)  compared to \textit{Choelesky \& Volrend\_npl} (around 1\%). For other workload, this hit rate around 20\%. For  {write coalescing} (Figure \ref{fig:pb_hit_rate}(b)), we observe similar pattern except \textit{FFT} which also have lower coalescing rate (2.8\%). As \textit{Radiocity} have highest read hit rate and write coalescing rate, it experience higher increase of speedup in PB\_RF scheme over PB scheme. Where for \textit{Choelesky \& Volrend\_npl} very low read hit rate and write coalescing rate results in slowdown  {over PB scheme}. As \textit{Lu\_cont, Lu\_non \& Raytrace} have noticeable read hit rate and write coalescing rate, they also demonstrate speedup gain for PB\_RF over PB.

\begin{figure}[!ht]
    \centering
    \includegraphics[width=\linewidth]{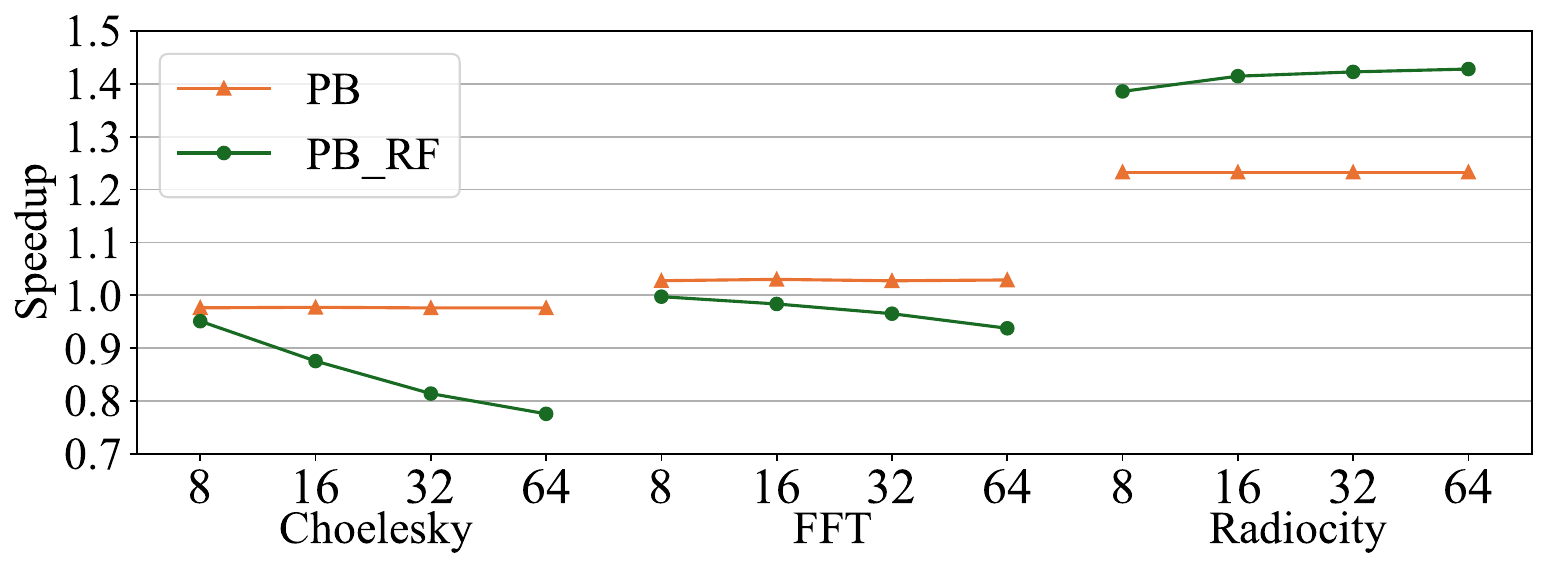}
    \caption{ Speedup of PB and PB\_RF with increasing PBE count in PCS.(higher is better). x-axis represent PBE count in the CXL switch.}
    \label{fig:pb_entry_sensitivity}
\end{figure}


Figure \ref{fig:pb_entry_sensitivity} shows speedup  of  \textit{Radiocity, Choelesky \& FFT}  over NoPB for different PBE count. \textit{Radiocity} experienced high speedup for both PB and PB\_RF, representing best case scenario while \textit{Choelesky} results in maximum slowdown for both representing worst case scenario. \textit{FFT} experience speedup for PB but slowdown for PB\_RF highlight contrasting  effect between two scheme. With PBE count increase, tag access latency also increase which we calculated using cacti\cite{cacti1996}. For PB,  increasing PBE count does not show noticeable influence over speedup. For PB\_RF scheme, both \textit{Choelesky \& FFT} experience  increasing slowdown  with increase of PBE count, while \textit{Choelesky}  experience at higher rate. \textit{Radiocity} experience increasing speedup with increasing PBE count.




\section{Conclusion}
In this paper, we argue the need for persistent CXL switch (PCS) in CXL attached persistent memory system. To demonstrate our idea, we present a persist buffer(PB) design for PCS. We discuss potential performance and correctness challenges to design PB in CXL switch and show that simple persistent buffer could reduce data persist latency significantly. We also discuss opportunity to reduce read latency by forwarding  data copy from PB. Our evaluation show that, regardless simplistic design of PB, PCS achieves 12\% average speedup and PB\_RF optimization scheme achieves 15\% average speedup in Splash-4 workload compared to volatile CXL switch system.



\bibliographystyle{unsrt}
\bibliography{references}

\end{document}